\documentclass[12pt]{article}
\begin{document}
\begin{center}
 {\bf Analytic proof of the Sutherland conjecture}\\
\vspace*{1cm} J. Dittrich\footnote{Also a member of the Doppler
Institute for Mathematical Physics, Faculty of Nuclear Sciences
and Physical Engineering, Czech Technical University, Prague.}
\\ {\small Nuclear Physics Institute ASCR, CZ-250 68 \v Re\v z, Czech Republic}\\ and\\
 V.I. Inozemtsev\\
{\small BLTP JINR, 141980 Dubna, Moscow Region, Russia}
\end{center}
\begin{tabbing}
\kill
\end{tabbing}
\begin{center}
{\sc Abstract\\}
\end{center}

Using the integral representation of the inverse of the
logarithmic derivative of the elliptic theta function, the
spectrum of the Lax matrix for the 1D system of particles
interacting via inverse sinh-squared potential is shown to be
given by the asymptotic Bethe ansatz
 in the thermodynamic limit.

\newpage

 The problem of the verification of the asymptotic Bethe ansatz method
[1] still remains unsolved more than 30 years after its first
 presentation. The method consists in using only scattering data for
the description of the integrable many-body systems in the
thermodynamic limit. It is well known that if the system is
integrable in the Yang-Baxter sense, the many-body scattering
matrix is expressed via only two-particle phase shift but real
structure of the wave functions might be rather complicated if the
interaction is nonlocal. In particular, when the two-body
potential is of the form $\sin^{-2}(\pi x/L)$, where $L$  is the
size of the system, or $1/\sinh^{2}(x)$, the wave functions differ
drastically from the linear combinations of the plane waves
inherent for the Bethe ansatz. Despite the exact results in the
thermodynamic limit available for $\sin^{-2}(\pi x/L)$ case are in
complete coincidence with the asymptotic Bethe ansatz, the {\it
reason} for this coincidence is still quite mysterious and it
cannot be used as an argument to validate the method for the case
of $1/\sinh^2 x$ pair potential which is more complicated from the
mathematical viewpoint .

Due to the lack of a general approach to the problem, any {\it
particular} exact results confirming the asymptotic Bethe ansatz
are of interest. Some years ago, Sutherland [2] proposed one
example of very good numerical coincidence of the asymptotic
results with exact ones. It concerns the densities of the
distribution of the eigenvalues of the $L$ matrix from the Lax
pair [3] for the systems of particles interacting via $1/\sinh^2
x$ potential,
$$L_{jk}=p_{j}\delta_{jk}+(1-\delta_{jk})i\lambda\coth(x_{j}-x_{k}),$$
where $p_{j}=-i\partial/\partial x_{j}$ obey the canonical
commutation relations $[x_{j},p_{k}]=i\delta_{jk}.$ The
corresponding Hamilton operator reads
$$H={1\over2}\sum_{j=1}^{N}p_{j}^2
+\lambda^2\sum_{j<k} {1\over
{\sinh^2(x_{j}-x_{k})}}.$$ Asymptotically, if $x_{1}\ll...\ll
x_{N}$, the particles have the momenta $k_{1}<...<k_{N}$ and the
elements of the Lax matrix become $c$-numbers,
$$(L_{as})_{jk}=k_{j}\delta_{jk}+i\lambda\,{\rm
sgn}(j-k).\eqno(1)$$ The asymptotic Bethe ansatz gives the
asymptotic momenta as solutions to the equations $$Lk_{j}=2\pi
I_{j}+\sum_{l\neq j}^{N}\tau(k_{j}-k_{l}), \eqno(2)$$ where
$\tau(k)$ is the two-body phase shift, $L$ is the total size of a
 system and $\{I_{j}\}$ are quantum numbers. In the classical limit
$\lambda\to \infty$, for the ground state, (2) becomes an integral equation
for large $N$ and $L$ [4],
$$ 2a=\int_{-A}^{A}dx'\gamma(x-x')\rho(x'),\eqno(3)$$
where
$$\gamma(x)=\ln (1+x^{-2}),$$
$\rho(x)$ is the density of the momentum distribution in the ground state,
and $a=L/N$ is average nearest-neighbor spacing (or the lattice constant).
The kernel of this integral equation is symmetric and positive definite.
Thus it has unique solution at given $A$ [6]. The normalization condition
$$\int _{-A}^{A}\rho(x')dx'=1\eqno(4)$$
defines $A$ as a function of $a$.
The distribution of the eigenvalues of the Lax matrix can be connected with
the distribution  of the momenta [2]. Indeed, it follows from (1) that the
equation for the eigenvalues
$${\rm det}(L_{as}-Iz)=
{1\over2}\left[\prod_{j=1}^{N}(k_{j}-z+i\lambda)+
\prod_{j=1}^{N}(k_{j}-z-i\lambda)\right]
 $$
$$=\prod_{s=1}^{N}(\omega_{s}-z)=0$$
can be written as
$$(s+1/2)\pi={1\over{2i}}\sum_{j=1}^{N
}{\rm ln}\left[{{k_{j}-\omega_{s}+i\lambda}\over
{k_{j}-\omega_{s}-i\lambda}}\right]=\sum_{j=1}^{N}
{\rm arctan}\left[{\lambda\over{k_{j}-\omega_{s}}}\right].$$
A discontinuous branch of arctan with values in $[0,\pi]$ is used
here.
In the thermodynamic limit, the eigenvalues $\{\omega\}$ are
distributed with the density $\sigma(\omega)$:
$N\sigma(\omega)d\omega$ gives the number of $\{\omega\}$ in the interval
$(\omega,\omega+d\omega)$. Hence
$${{ds}\over{d\omega}}=N\sigma(\omega).$$
Differentiating the above relation with respect to $\omega$ and
taking classical limit gives
$$\sigma(\omega)={1\over{ 2\pi}}\int_{-A}^{A}{{dx\rho(x)}\over
{(x-\omega)^2+1/4}}\eqno(5)$$
after rescaling of variables [2].
One can see that in the classical limit the density $\sigma(\omega)$
can be calculated via the solution of the integral equation of the
asymptotic Bethe ansatz method (3).

On the other hand, in this limit the particles take their equilibrium
 positions at $x_{j}=ja$ in the ground state with $k_{j}=0$. The form of
the Lax matrix becomes simpler,
$$L_{jk}=(1-\delta_{jk})\,i\,\lambda\,{\rm coth}[a(j-k)],\eqno(6)$$
and the distribution of its eigenvalues can be calculated directly since
(6) is of the Toeplitz form. Its eigenvectors are the plane waves, and after
imposing periodic boundary condition (i.e. regularization of the determinant)
and taking thermodynamic limit $N\to\infty$ one could introduce the variable
$\phi=2\pi s/N$ defining the continuous distribution
$$\omega(\phi)=\omega_{s}=\omega_{N\phi/2\pi}.$$
 The result can be written upon rescaling
 $\omega \to 2\lambda\omega$
in the form  [2]
$$\omega(\phi)=-{{\theta'_{1}(\phi/2)}\over{2\theta_{1}(\phi/2)}},\eqno(7)$$
where $\theta_{1}(x)$ is the standard theta function
$$\theta_{1}(x)=2\sum_{n=0}^{\infty}(-1)^{n}q^{(n+1/2)^2}\sin (2n+1)x$$
 with the nome $ q=e^{-a}$.
The density of the eigenvalues is given now by the formula
$$\sigma(\omega)={1\over{ 2\pi}}{{d\phi}\over{d\omega}},\eqno(8)$$
where the derivative is calculated through the relation
$${{d\omega}\over{d\phi}}=-{1\over 4}
\left[{{\theta'_{1}(\phi/2)}\over{\theta_{1}(\phi/2)}}\right]'$$
$$=-{{K^2}\over{\pi^2}}\left[{{K-E}\over{K}}-{1\over{{\rm sn}^2(K\phi/\pi)}}
\right].\eqno(9)$$
Thus one gets two representations for the density of the eigenvalues
of the classical Lax matrix, one exact ( formulae (7-9)) and one obtained
by using the asymptotic Bethe ansatz method (formulae (3-5)). If it is true,
 they should
coincide. The main difficulty in verifying this fact is that there
is no chance to find analytic solution to the integral equation
(3). In [2], Sutherland found good coincidence of both expressions
by solving this equation numerically with high accuracy. However,
analytic solution of the problem has not been found.

In what follows, we propose a construction which uses analytic properties of
the elliptic functions and provides the desired proof.
Let us introduce the notation $\chi_r=\Re e \chi$, $\chi_i=\Im m
\chi$ for any complex $\chi$.
Consider at first the problem of explicit construction of the inverse function
$\phi(\omega)$ such that
$$\phi(\omega(\lambda))\equiv\lambda.$$
It is clear that it is no longer holomorphic in the $\omega$-plane. Indeed,
on the lines $\phi=\phi_{r}\pm ia$ one finds $\omega(\phi)=\omega_{r}\pm
i/2$ due to the quasiperiodicity property
$$\omega(\phi+2ia)=\omega(\phi)+i, \quad \omega(\phi+2\pi)=\omega(\phi).$$
One has also
$$\omega_{r}(\pm ia)=\omega_{r}(2\pi\pm ia)=0.$$
The derivative ${{d\omega}\over{d\phi}}$ is double periodic with periods
$(2\pi, 2ia)$ and has double pole in the fundamental domain
$$0\leq \Re e\phi<2\pi,\quad -a\leq \Im m \phi<a.$$
Therefore it must have just two zeros giving two extremal points of
 $\omega(\phi)$:
one minimum of $\omega_{r}$ located at $\phi_{min}+ia$ and one maximum
located at $2\pi-\phi_{min}+ia$.
Both these extrema are considered with respect to line
$\Im m\,\phi = a$, and $\phi_{min}\in (0,2\pi)$.
Let us denote $\Omega_{0}=\omega_{r}(2\pi-\phi_{min}+ia)$. Then   it is
evident that the function $\phi(\omega)$ should have two cuts in the
 $\omega$-plane represented by the segments $-\Omega_{0}\leq \omega_{r}
\leq \Omega_{0},$ $\omega_{i}=1/2$ and $-\Omega_{0}\leq \omega_{r}\leq
\Omega_{0}$, $\omega_{i}=-1/2$. Following Haldane [5], let us express $\phi$ as a
Cauchy integral over the contour along the image of the boundary of the fundamental
domain and use the symmetry properties of $\omega(\phi)$. We skip these rather
long but in fact simple considerations. Only integral over the finite interval
remains and after integrating by parts we obtain
$$\phi(\omega)=\int_{-\Omega_{0}}^{\Omega_{0}}dx\rho_{0}(x){1\over i}
\ln {{\omega-x-i/2}\over{\omega-x+i/2}}.\eqno(10)$$
The still unknown function $\rho_0(x)$ is normalized due to the properties of
the function $\phi$: $\omega(\phi+2\pi)=\omega(\phi)$ and the
integral representation (10),
$$\int_{-\Omega_{0}}^{\Omega_{0}}\rho_{0}(x)dx=1.\eqno(11)$$
On the other hand, we know that $\phi_{i}(\omega\pm i/2)=\pm ia$ for all real
$\omega$ in the interval $-\Omega_{0}\leq \omega\leq \Omega_{0}$.
This gives an integral equation for the function $\rho_{0}(x)$ entering
(10) of the form quite similar to (3),
$$\Im m\phi(\omega+i/2)=a={1\over2}\int_{-\Omega_{0}}^{\Omega_{0}}
dx\rho_{0}(x)\ln (1+(\omega-x)^{-2}).\eqno(12)$$
Note also that the same equation can be obtained with the use of
quasiperiodicity property $\omega(\phi+2ia)=\omega(\phi)+i$ and the
representation (10).

The equations (3) and (12) become completely identical  if one puts
$A=\Omega_{0}$, i.e. the meaning of the parameter $A$ is that it defines
 the maximal value of $\omega_{r}(\phi)$ on the segment $0\leq \phi_{r}\leq
2\pi,$ $\phi_{i}=ia$ due to the uniqueness of the solution of (3) mentioned
 above.

It is straightforward now to verify by differentiating  (10) with
respect to $\omega$ that the derivative ${{d\omega}\over{d\phi}}$
has the integral representation
$$\left({{d\omega}\over{d\phi}}
\right)^{-1}=\int_{-\Omega_{0}}^{\Omega_{0}}
{{dx\rho_{0}(x)}\over{(\omega-x)^2+1/4}}.\eqno(13)  $$
Comparing
both sets of formulas (3-5) and (11-13), one can easily see that
the expressions for the spectral density of the Lax matrix in the
classical limit coincide after identification
$\rho(x)=\rho_{0}(x)$. This completes the proof.
\\ \\
{\bf Acknowledgements}. On of us (V.I.) is grateful to Prof.
F.D.M. Haldane for giving him a copy of the paper [5]. The work
was supported in part by the Votruba-Blokhintsev program and ASCR
grant no. IAA1048101.
\newpage

{\bf References}\\

\begin{enumerate}
\item
B. Sutherland. J.Math.Phys. 12, 251 (1971)
\item
B. Sutherland. Phys. Rev. Lett. 75, 1248 (1995)
\item
F. Calogero, C. Marchioro and O. Ragnisco. Lett. Nuovo Cim. 13, 383 (1975)
\item
B. Sutherland, R.A. R\"omer and B.S. Shastry. Phys. Rev. Lett. 73, 2154 (1994)
\item
F.D.M. Haldane (unpublished)
\item
E.T. Whittaker and G.N. Watson. A Course of Modern Analysis, Cambridge
at the University Press, 1927
\end{enumerate}
\end{document}